\documentclass{aa}

\usepackage{txfonts}
\usepackage{graphicx}

\begin{document}

\title{Nonlinear Fast Magnetoacoustic Wave Propagation \\ in the Neighbourhood of a 2D magnetic X-point: \\ Oscillatory Reconnection}

\author{J.~A.~McLaughlin\inst{1} \and I.~De Moortel\inst{1} \and A.~W.~Hood\inst{1} \and C.~S.~Brady\inst{2}}

\offprints{J.~A.~McLaughlin, \email{james@mcs.st-and.ac.uk}}

\institute{School of Mathematics and Statistics, University of St
Andrews, KY16 9SS, UK \and Physics Department, University of Warwick, CV4 7AL, Coventry, UK}

\date{Received 26 June 2008 / Accepted 15 October 2008}

\authorrunning{McLaughlin {{et al.}}}
\titlerunning{Nonlinear fast wave propagation around an X-point}

\abstract{This paper  extends the models of Craig \& McClymont (1991) and McLaughlin \& Hood (2004) to include finite $\beta$ and nonlinear effects.}
{We investigate the nature of nonlinear fast magnetoacoustic waves about a 2D magnetic X-point.}
{We solve the compressible and resistive MHD equations using a Lagrangian remap, shock capturing code (Arber et al. 2001) and consider an initial condition in $ {\bf{v}}\times{\bf{B}} \cdot {\hat{\bf{z}}}$ (a natural variable of the system).}
{We observe the formation of both fast and slow oblique magnetic shocks. The nonlinear wave deforms the X-point into a \lq{cusp-like}\rq{} point which in turn collapses to a current sheet. The system then evolves through a series of horizontal and vertical current sheets, with associated changes in connectivity, i.e. the system exhibits oscillatory reconnection. Our final state is non-potential (but in force balance) due to asymmetric heating from the shocks. Larger amplitudes in our initial condition correspond to larger values of the final current density left in the system.}
{The inclusion of nonlinear terms introduces several new features to the system that were absent from the linear regime.}

\keywords{Magnetohydrodynamics (MHD) -- Waves -- Shock waves -- Sun:~corona -- Sun:~ magnetic fields -- Sun:~oscillations    }

\maketitle


\section{Introduction}\label{section1}

It is now known that MHD wave motions (e.g.  Roberts \cite{Bernie}; De Moortel \cite{DeMoortel2005};  Nakariakov \& Verwichte \cite{NV2005}) are omnipresent throughout the solar corona (Tomczyk et al. \cite{Tomczyk}). Many solar instruments have observed various MHD wave motions in the solar atmosphere: slow magnetoacoustic (MA) waves have been seen in {\emph{SOHO}} data (Berghmans \& Clette \cite{Berghmans1999}; Kliem et al. \cite{Kliem}; Wang et al. \cite{Wang2002}) and {\emph{TRACE}} data (De Moortel et al. \cite{DeMoortel2000}). Fast MA waves have been seen with {\emph{TRACE}} (Aschwanden et al. \cite{Aschwandenetal1999}, \cite{Aschwandenetal2002}; Nakariakov et al. \cite{Nakariakov1999}; Wang \& Solanki \cite{Wang2004}) and {\emph{Hinode}} (Ofman \& Wang \cite{OW2008}). Non-thermal line narrowing / broadening due to Alfv\'en waves has been reported by Harrison et al. (\cite{Harrison2002}) / {Erd{\'e}lyi} et al. (\cite{E1998}) and  O'Shea et al. (\cite{Oshea}) . More recently, Alfv\'en waves have been observed in the corona  (Okamoto et al. \cite{Okamoto}; Tomczyk et al. \cite{Tomczyk}) and chromosphere (De Pontieu et al. \cite{Bart2007}), although these claims are currently subject to intense discussion (Erd{\'e}lyi \& Fedun \cite{RF2007}; Van Doorsselaere et al. \cite{Tom2008}).

It is clear that the  coronal magnetic field  plays a fundamental role in the  propagation and properties of MHD waves, and to begin to understand this  inhomogeneous, magnetised  environment, it is useful to look at the topology (structure) of the magnetic field itself.  Potential-field extrapolations of the coronal magnetic field can be made from photospheric magnetograms, and such extrapolations show the existence of important features of the topology: {\it{null points}} - locations in the field where the magnetic field, and hence the Alfv\'en speed, is zero, and {\it{separatrices}} - topological features that separate regions of different magnetic flux connectivity. Detailed investigations of the coronal magnetic field, using such potential field calculations, can be found in e.g.  Brown \& Priest (\cite{Brown2001}),  Beveridge et al. (\cite{Beveridge2002}), R{\'e}gnier et al. (\cite{RPH2008}) or a more comprehensive review  by Longcope (\cite{L2005}).

The propagation of fast magnetoacoustic waves in an inhomogeneous coronal plasma has been investigated by Nakariakov \& Roberts (\cite{Nakariakov1995}), who showed that the waves are refracted into regions of low Alfv\'en speed. In the case of null points, the Alfv\'en speed actually drops to zero.

McLaughlin \& Hood (2004) (hereafter referred to as {\cite{MH2004}}) solved the linearised, $\beta=0$  MHD equations using a two-step Lax-Wendroff numerical scheme. They found that in the neighbourhood of a single 2D X-point, the fast MA wave refracted around and accumulated at the null point. These key results have been found to carry over from the simple 2D single magnetic null point to two null points (McLaughlin \& Hood \cite{MH2005}) and to a more realistic magnetic configuration of a null point created by two dipoles (McLaughlin \& Hood \cite{MH2006a}). It should be noted that the behaviour of the fast wave is entirely dominated by the Alfv\'en-speed profile, and since the magnetic field drops to zero at the null point, the wave will never reach the actual null for a $\beta=0$ plasma. McLaughlin \& Hood (\cite{MH2006b}) extended the model of {\cite{MH2004}} to include plasma pressure effects. This naturally led to the inclusion of the slow MA wave  and the introduction of a $\beta=1$ layer around the null point; representing a high $\beta$ environment inside and low outside. Coupling and mode conversion is observed at locations where the sound speed and Alfv\'en speed become comparable in magnitude (e.g. Zhugzhda \& Dzhalilov \cite{ZD1982}; Cally \cite{Cally}; Bloomfield et al. \cite{Bloomfield}; McDougall \& Hood \cite{Dee}).

Waves in the neighbourhood of a single 2D null point have been investigated by various authors. Bulanov \& Syrovatskii (\cite{Bulanov1980}) provided a detailed discussion of the propagation of harmonic fast and Alfv\'en waves using cylindrical symmetry. Craig \& Watson (\cite{CraigWatson1992}) mainly considered the radial propagation of the $m=0$ mode (where $m$ is the azimuthal wavenumber) using a mixture of analytical and numerical solutions. They showed that the propagation of the $m=0$ wave towards the null point generates an exponentially large increase in the current density and that magnetic resistivity dissipates this current in a time related to $\log { \eta }$. Craig \& McClymont (\cite{CraigMcClymont1991, CraigMcClymont1993}), Hassam (\cite{Hassam1992}) and Ofman et al. (\cite{OMS1993}) investigated the normal mode solutions for both $m=0$ and $m\ne 0$ modes with resistivity included. Again, they emphasise that the current builds up as the inverse square of the radial distance from the null point. All these investigations were carried out using cylindrical models in which the generated waves encircled the null point.

Reconnection can occur when strong currents cause the magnetic fieldlines to diffuse through the plasma and change their connectivity (Parker \cite{Parker}; Sweet \cite{Sweet}; Petschek \cite{Petschek}). In 2D, reconnection can only occur at null points (Priest \& Forbes \cite{magneticreconnection2000}). Dungey (\cite{Dungey}) reported that a perturbed X-point can collapse if the footpoints of the field are free to move, Mellor et al. (\cite{Mellor}) looked at the linear collapse of a 2D null point, and Imshennik \& Syrovatsky (\cite{IS1967}) described the collapse with an exact, non-linear solution of the ideal MHD equations. However, these papers did not include the effect of gas pressure, which would act to limit the growth of the current density.  In considering the relaxation of a 2D X-type neutral point disturbed from equilibrium,   Craig \& McClymont (\cite{CraigMcClymont1991}) found that free magnetic energy is dissipated by {\emph{oscillatory reconnection}}, which couples resistive diffusion at the null to global advection of the outer field. An example of oscillatory reconnection generated by flux emergence within a coronal hole was recently detailed by Murray et al. (\cite{Murray}). Finally, Longcope \& Priest (\cite{LP2007}) investigated the diffusion of a 2D current sheet  subject to suddenly enhanced resistivity. They found that the diffusion couples to a fast MA mode which propagates the current  away at the local Alfv\'en speed.

The aim of this paper is to extend the model used in {\cite{MH2004}} and Craig \& McClymont (\cite{CraigMcClymont1991}) to include finite $\beta$ and nonlinear effects. To realise this, we will be solving the compressible and resistive MHD equations using a Lagrangian remap, shock capturing code: {\emph{LARE2D}} (Arber et al. \cite{Arber}). The key results from {\cite{MH2004}}, i.e. for the linear fast wave, are that it demonstrates refraction, that the wave energy accumulates at the null, and that current density builds up exponentially at that point. We believe it is important to extend this work to include nonlinear effects since {\cite{MH2004}} indicated a preferential topological location for (ohmic) heating and we need to see if this observational prediction persists when we consider larger (and hence nonlinear) wave amplitudes. This paper will address three main questions that naturally arise when extending {\cite{MH2004}} into the nonlinear regime:
\begin{itemize}
\item[(1)]{Does the fast wave now steepen to form shocks, and can these propagate across or escape the null?}
\item[(2)]{Can the refraction effect drag enough magnetic field into the null to initiate X-point collapse or reconnection?}
\item[(3)]{Has the rate of current density accumulation changed, and is the null still the preferential location of wave heating?}
\end{itemize}

The paper has the following outline: the basic setup, equations and assumptions are described in \S\ref{section2}, \S\ref{section3}  details the nonlinear fast MA wave behaviour  and the resultant oscillatory reconnection is discussed in \S\ref{section4}. We consider different amplitudes for our initial condition in \S\ref{section5} and the conclusions are given in \S\ref{section:conclusions}.


\begin{figure}[t]
\begin{center}
\includegraphics[width=4.0in]{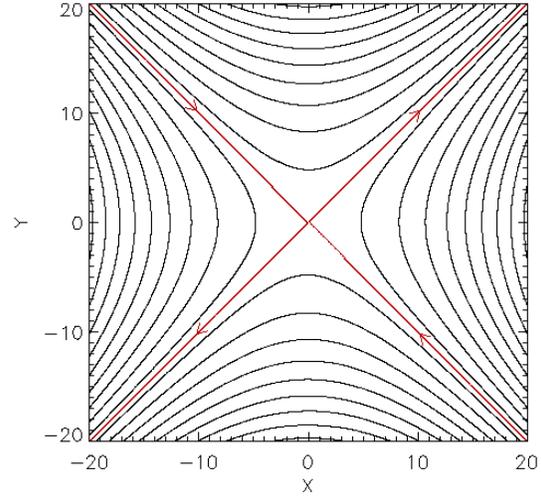}
\caption{The equilibrium magnetic field. The red lines denote the separatrices and arrows indicate the direction of the magnetic field. The null point is located at the origin, where the separatrices intersect.}
\label{figureone}
\end{center}
\end{figure}

\section{Basic Equations}\label{section2}

We consider the 2D compressible and resistive MHD equations appropriate to the solar corona:
\begin{eqnarray}
\qquad \rho \left[ {\partial {\bf{v}}\over \partial t} + \left( {\bf{v}}\cdot\nabla \right) {\bf{v}} \right] &=& - \nabla p + \left( {{\frac{1}{\mu}}}   \nabla \times {\bf{B}}  \right)\times {\bf{B}}      \; ,\nonumber \\
 {\partial {\bf{B}}\over \partial t}  &=& \nabla \times \left ({\bf{v}}\times {\bf{B}}\right ) + \eta \nabla ^2  {\bf{B}}\; ,\nonumber \\
{\partial \rho\over \partial t} + \nabla \cdot \left (\rho {\bf{v}}\right ) &=& 0\; , \nonumber \\
 \rho \left[{\partial {\epsilon}\over \partial t}  + \left( {\bf{v}}\cdot\nabla \right) {\epsilon}\right] &=& - p \nabla \cdot {\bf{v}} + {{\frac{1}{\sigma}}} \left| {\bf{j}} \right| ^2  \; \label{MHDequations}  ,
\end{eqnarray}

where $\rho$ is the mass density, ${\bf{v}}$ is the plasma velocity, ${\bf{B}}$ the magnetic induction (usually called the magnetic field), $p$ is the plasma pressure,  $ \mu = 4 \pi \times 10^{-7} \/\mathrm{Hm^{-1}}$  is the magnetic permeability, $\sigma$ is the electrical conductivity,  $\eta=1/ {\mu \sigma} $ is the magnetic diffusivity, $\epsilon= {p / \rho \left( \gamma -1 \right)}$ is the specific internal energy density, where $\gamma={5 / 3}$ is the ratio of specific heats and ${\bf{j}} = {{\nabla \times {\bf{B}}} / \mu}$ is the electric current density.

The {\emph{LARE2D}} numerical code utilises artificial shock viscosity to introduce dissipation at steep gradients. The details of this technique, often called Wilkins  viscosity, can be found in Wilkins (\cite{Wilkins1980}) and Arber et al. (\cite{Arber}).

\newpage

\subsection{Basic equilibrium  and non-dimensionalisation}\label{section:2.1}

\begin{figure*}
\begin{center}
\includegraphics[width=7.5in]{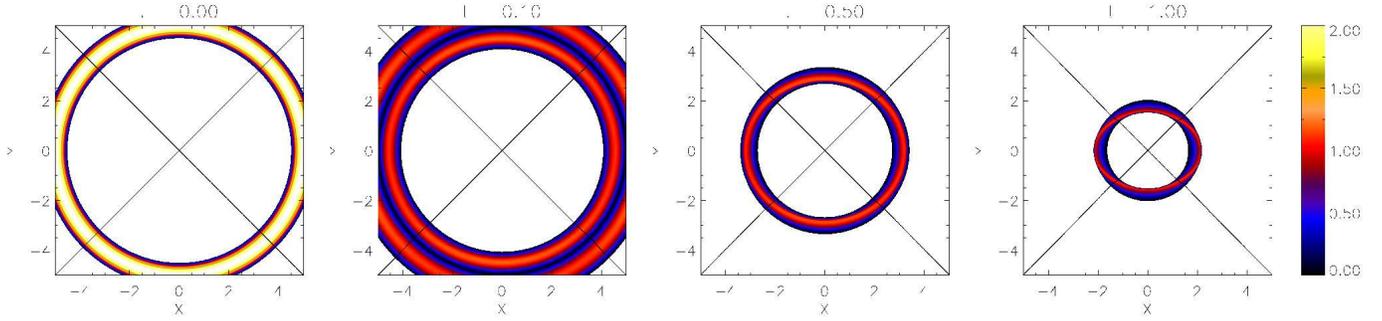}
\caption{Contours of ${\rm{v}}_\perp$ for a fast wave pulse initially located at a radius $r=5$, and its resultant propagation at (Alfv\'en) times $t=0, 0.1,0.5 \;\& \;1$. The black lines denote the separatrices and the null point is located at their intersection (the origin).  The full evolution, $0\le t\le60$, is available as an mpeg animation in the  online edition of the Astropnomy \& Astrophysics Journal.}
\label{figuretwo}
\end{center}
\end{figure*}

The equilibrium magnetic field structure is taken as a simple 2D X-type neutral point. The aim of studying waves in a 2D configuration is one of simplicity. There are many complicated effects including mode conversion and coupling, and a 2D geometry allows us to understand and explain these behaviours better, before the extension to 3D. The initial magnetic field is taken as
\begin{eqnarray}
\qquad {\bf{B}}_0 = \frac{B}{L} \left(y, x, 0\right) \;,   \label{Xpoint}
\end{eqnarray}
where $B$ is a characteristic field strength and $L$ is the length scale for magnetic field variations. This magnetic field can be seen in Figure \ref{figureone}. Equation (\ref{Xpoint}) is slightly different to that used in {\cite{MH2004}}, where the authors considered ${\bf{B}}_0=\left(x,0,-z\right)$. This simply represents a $\pi / 4$ rotation of our magnetic field and the key results are still valid. Note that this magnetic configuration is no longer valid far from the null point, as the field strength tends to infinity. However, McLaughlin \& Hood (\cite{MH2006a}) looked at a $\beta=0$ magnetic field that decays far from the null and they found that the key results from \cite{MH2004} remain true close to the null.

We will also find it useful to consider ${\bf{A}}$, the vector potential, where ${\bf{B}}= \nabla \times {\bf{A}}$. In 2D and for the coordinate system used in this paper, ${\bf{A}}=A_z {\hat{\bf{z}}}$ and our equilibrium vector potential is given by:
\begin{eqnarray}
\qquad {\bf{A}}_0 = A_0{\hat{\bf{z}}}  =  \frac{1}{2}\left( y^2-x^2\right) {\hat{\bf{z}}} \;.  \label{vectorpotential}
\end{eqnarray}

We now consider a change of scale to non-dimensionalise all variables. Let ${\rm{\bf{v}}} = {\rm{v}}_0 {\mathbf{v}}^*$,  ${\mathbf{B}} = B {\mathbf{B}}^*$, $x = L x^*$, $y=L y^*$, $\rho={\rho}_0 \rho^*$, $p = p_0 p^*$, $\nabla = \frac{1}{L}\nabla^*$, $t={t}_0 t^*$, ${\bf{A}}=B L {\bf{A}}^*$ and $\eta = \eta_0$, where we let * denote a dimensionless quantity and ${\rm{v}}_0$, $B$, $L$, ${\rho}_0$, $p_0$, ${t}_0$ and $\eta_0$ are constants with the dimensions of the variable they are scaling. We then set $ {B} / {\sqrt{\mu \rho _0 } } ={\rm{v}}_0$ and ${\rm{v}}_0 =  {L} / {{t_0}}$ (this sets ${\rm{v}}_0$ as a constant background Alfv\'{e}n speed). We also set ${\eta_0 {t}_0 } /  {L^2} =R_m^{-1}$, where $R_m$ is the magnetic Reynolds number, and set $ {\beta_0} = {2 \mu p_0} / {B^2}$, where $\beta_0$ is the plasma-$\beta$ at a radius $L$ from the origin. This process non-dimensionalises equations (\ref{MHDequations}) and under these scalings, $t^*=1$ (for example) refers to $t={t}_0=  {L} / {{\rm{v}}_0}$; i.e. the time taken to travel a distance $L$ at the background Alfv\'en speed. For the rest of this paper, we drop the star indices; the fact that all variables are now non-dimensionalised is understood.

We take the equilibrium density to be uniform, i.e. $\rho=\rho_0$;  a spatial variation in $\rho_0$ can cause phase mixing (Heyvaerts \& Priest \cite{Heyvaerts1983}; De Moortel {et al.} \cite{DeMoortel1999}; Hood et al. \cite{Hood2002}). We set $R_m=10^4$.

Finally, we consider the equilibrium plasma to be cold: $T=0$K  (i.e. $\beta_0=0$) and, hence, ignore plasma pressure effects (as in {\cite{MH2004}}). However, as we will see below, magnetic shocks heat the plasma  and so the plasma will not remain cold (see e.g. $\S 1.5$ in Priest \& Forbes \cite{magneticreconnection2000}).

\subsection{Initial and boundary conditions}\label{section:2.2}

Equations (\ref{MHDequations}) are solved numerically using a Lagrangian remap, shock-capturing code called {\emph{LARE2D}} (Arber et al. \cite{Arber}). The equations are solved computationally in a square domain $x,y   \in [-20,20]$ with a numerical resolution of $5120 \times 5120$.  Zero gradient boundary conditions are applied to the variables ${\bf{B}}$, $\rho$, $\epsilon$ at the four boundaries, and  ${\bf{v}}$ is set to zero on all boundaries, i.e. reflective boundaries. A damping region exists for $x^2+y^2 \ge 6$ and so all oscillations that enter this region are slowly damped away. The  (equilibrium) Alfv\'en speed increases with distance from the null point and, hence, waves accelerate as they propagate outwards. Since we do not want reflected waves to influence our null point, implementation of such a damping region is essential.

As seen in \cite{MH2004}, there are two natural variables to consider in our system: ${\rm{v}}_\perp= \left( {\bf{v}}\times{\bf{B}}\right) \cdot {\hat{\bf{z}}}={\rm{v}}_x B_y - {\rm{v}}_y B_x$ and  ${\rm{v}}_\parallel=  {\bf{v}}\cdot{\bf{B}}={\rm{v}}_x B_x + {\rm{v}}_y B_y$. Here, ${\rm{v}}_\perp$ and ${\rm{v}}_\parallel$ are related to the perpendicular and parallel velocity, respectively and, as seen in \cite{MH2004}, their implementation naturally simplifies the governing equations, aids in MHD mode interpretation  and  (for ${\rm{v}}_\perp$) led to an analytical solution to the linear, cold plasma equations.

In cartesian coordinates: ${\rm{v}}_x = \left( {\rm{v}}_\parallel B_x + {\rm{v}}_\perp B_y \right) / \left| {\bf{B}}\right|^2$ and ${\rm{v}}_y = \left( {\rm{v}}_\parallel B_y - {\rm{v}}_\perp B_x \right) / \left| {\bf{B}}\right|^2$. In polar coordinates: ${\rm{v}}_x = {\rm{v}}_\perp \cos{\theta} / r$, ${\rm{v}}_y = -{\rm{v}}_\perp \sin{\theta} / r$, where $r=\sqrt{x^2+y^2}$ and we take ${\rm{v}}_\parallel$ to be initially zero. We note that with our choice of magnetic null point (equation \ref{Xpoint}), if we drive any of the velocity variables, the system will naturally develop a $\theta$ dependence.

Paper I  clearly demonstrates that the Alfv\'en speed (${\rm{v}}_A^2=B_x^2+B_y^2=x^2+y^2$) plays a vital role. Hence, it is natural to consider either a polar coordinate system or a circular pulse. In addition, as commented by  McClements et al. (\cite{McClements}),  a disturbance initially consisting of a plane wave is
refracted as it approaches the null in such a way that it becomes more azimuthally symmetric. Thus, it is appropriate to consider the evolution of azimuthally symmetric perturbations.  We expect the nonlinear behaviour to be more complicated than the linear equivalent and so, in order to clearly demonstrate the differences between the two systems, in this paper we consider an initial condition in velocity, such that:
\begin{eqnarray}
{\rm{v}}_\perp \left(x,y,t=0 \right)  &=& 2C \sin \left[ \pi \left(r - 4.5 \right) \right]{\rm{\;\; for\;\;\;}}4.5\le r \le 5.5 \label{ICs}\;,\\
{\rm{v}}_\parallel \left(x,y,t=0 \right)  &=& 0\nonumber
\end{eqnarray}
where $2C$ is our initial amplitude. Initial condition (\ref{ICs}) describes a circular, sinusoidal pulse in ${\rm{v}}_\perp$.   When the simulation begins, this initial pulse will naturally split into two waves, each of amplitude $C$,  travelling in different directions: an outgoing wave and an incoming wave. In this paper we will focus on the incoming wave, i.e. the wave travelling towards the null point. The damping regions will remove kinetic energy from the outgoing waves, and so they do not influence the null.  The initial condition produces a propagating disturbance that crosses magnetic fieldlines. Hence, we identify these waves as (initially) fast MA waves.

By choosing a small value for $C$ in equation (\ref{ICs}), we can recover the linear results from {\cite{MH2004}}. This can be seen in Appendix \ref{AppendixA} where we set $C=0.001$. For the nonlinear work described in \S\ref{section3} and \S\ref{section4}, we set $C=1$. Different values of $C$ will be considered in \S\ref{section5}.

\newpage

\section{Nonlinear fast MA behaviour}\label{section3}

\begin{figure}[t]
\begin{center}
\includegraphics[width=3.75in]{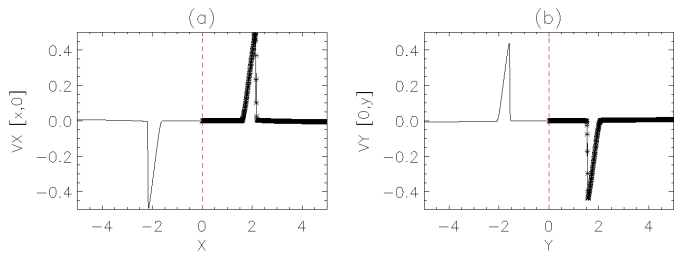}
\caption{$(a)$ Plot of ${\rm{v}}_x(x,0)$ for $-5\le x \le 5$ at $t=1$. $(b)$ Plot of ${\rm{v}}_y(0,y)$ for $-5\le y \le 5$ at $t=1$. The red line indicates the location of the null point. For $x,y \ge 0$, we have plotted stars to indicate the grid resolution.}
\label{figurethree}
\end{center}
\end{figure}

\subsection{Development of shocks $(0\le t\le1)$}\label{section3.1}

The evolution of  ${\rm{v}}_\perp$ can be seen in Figure \ref{figuretwo} ($0\le t\le1$), Figure \ref{figurefive} ($1.4\le t \le 2.8$) and Figure \ref{figurenine} ($2.9 \le t \le 60$), and the full evolution, $0\le t\le60$, is available as an mpeg animation in the  online edition of the Journal.

From Figure  \ref{figuretwo}, we see that the  initial pulse has split into two oppositely travelling wave pulses, where both waves can be seen at $t=0.1$. In order to clearly show the wave behaviour, we have plotted $x,y \in [-5,5]$, i.e. a square subsection of our total computational box. Hence, only the incoming wave can be seen after time $t=0.1$. This figure can be compared with Figure \ref{figureappendixA}, which corresponds to the linear regime.

There are two key features to note from Figure  \ref{figuretwo}. Firstly, we see that the incoming wave propagates across the magnetic fieldlines and keeps its initial pulse profile, i.e. an annulus, and the maximum amplitude remains at $C=1$.  The annulus contracts as the wave approaches the null point and this is the same refraction effect described in {\cite{MH2004}}. This refraction effect occurs since the (equilibrium) Alfv\'en speed is spatially varying.

Secondly, we note that the incoming wave pulse is developing an  asymmetry: the wave peaks are propagating faster (relative to the footpoints) in the $y-$direction than the $x-$direction. Thus, the wave pulse is developing discontinuities, where in the $y-$direction the wave peak is catching up with the leading footpoint, and in the $x-$direction the trailing footpoint is catching up with the wave peak. These discontinuities can be clearly seen in  Figure \ref{figurethree}a (trailing edge in ${\rm{v}}_x[x,0]$) and Figure \ref{figurethree}b (leading edge in ${\rm{v}}_y[0,y]$). We have also indicated the grid resolution (using stars) on the right-hand side of each subfigure which shows that the developing discontinuity is well resolved.

\begin{figure*}[t]
\begin{center}
\includegraphics[width=7.5in]{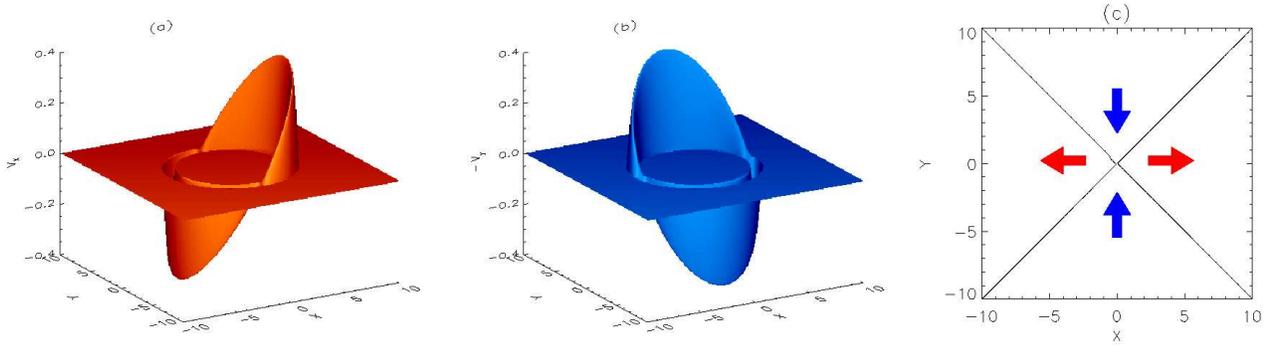}
\caption{Our choice of initial condition for $(a)$ ${\rm{v}}_x$  and $(b)$ ${\rm{v}}_y$ at $t=0$. This choice of initial condition prescribes a background velocity profile. $(c)$ Cartoon representation of the ${\rm{v}}_x$ (red arrows)  and ${\rm{v}}_y$ (blue) contributions to this background velocity profile, where the black lines denote the separatrices and the null point is located at their intersection.}
\label{figurefour}
\end{center}
\end{figure*}

This development of discontinuities was not reported in {\cite{MH2004}} as it is an entirely nonlinear effect, which arises because of our choice of a velocity initial condition (equation \ref{ICs}). In the nonlinear regime,  specifying an initial velocity condition  also prescribes a background velocity profile, and this profile can be seen in Figure \ref{figurefour}. It is this background velocity profile that leads to the development of discontinuities on the leading edges in the $y-$direction and on the trailing edges in the $x-$direction. The ${\rm{v}}_x$ (red arrows)  and ${\rm{v}}_y$ (blue arrows) contributions to this background velocity profile can be seen in Figure \ref{figurefour}c. This phenomenon is most easily understood by means of a simple 1D example, details of which can be found in Appendix \ref{AppendixB}. Note that the background velocity profile prescribed by equation (\ref{ICs}) appears as the $m=0$ mode in ${\rm{v}}_\perp$ but corresponds to the $m=2$ mode in cartesian components.

\newpage

\subsection{X-point collapse $(1.4\le t\le4)$}\label{section3.2}

\begin{figure*}[t]
\begin{center}
\includegraphics[width=7.5in]{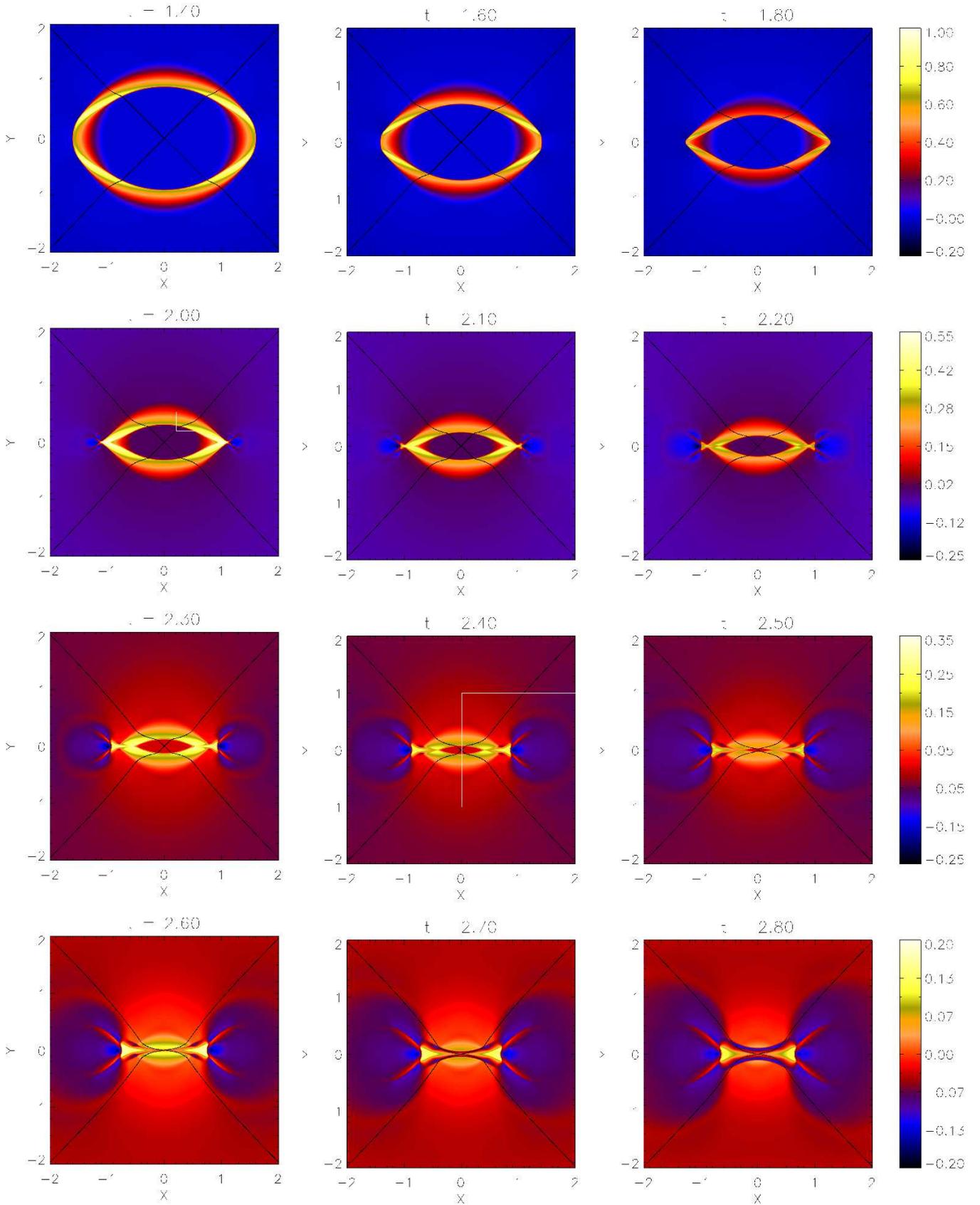}
\caption{Contours of ${\rm{v}}_\perp$ at various (Alfv\'en) times between $t=1.4$ and $t=2.8$. Note that the amplitude varies substantially throughout the evolution, and hence each row is assigned its own colour bar. The black lines denote the separatrices. The white boxes at $t=2$ \& $2.4$ are analysed in Figures \ref{figuresix} \& \ref{figureseven}, respectively.}
\label{figurefive}
\end{center}
\end{figure*}

\begin{figure*}[t]
\begin{center}
\includegraphics[width=7.5in]{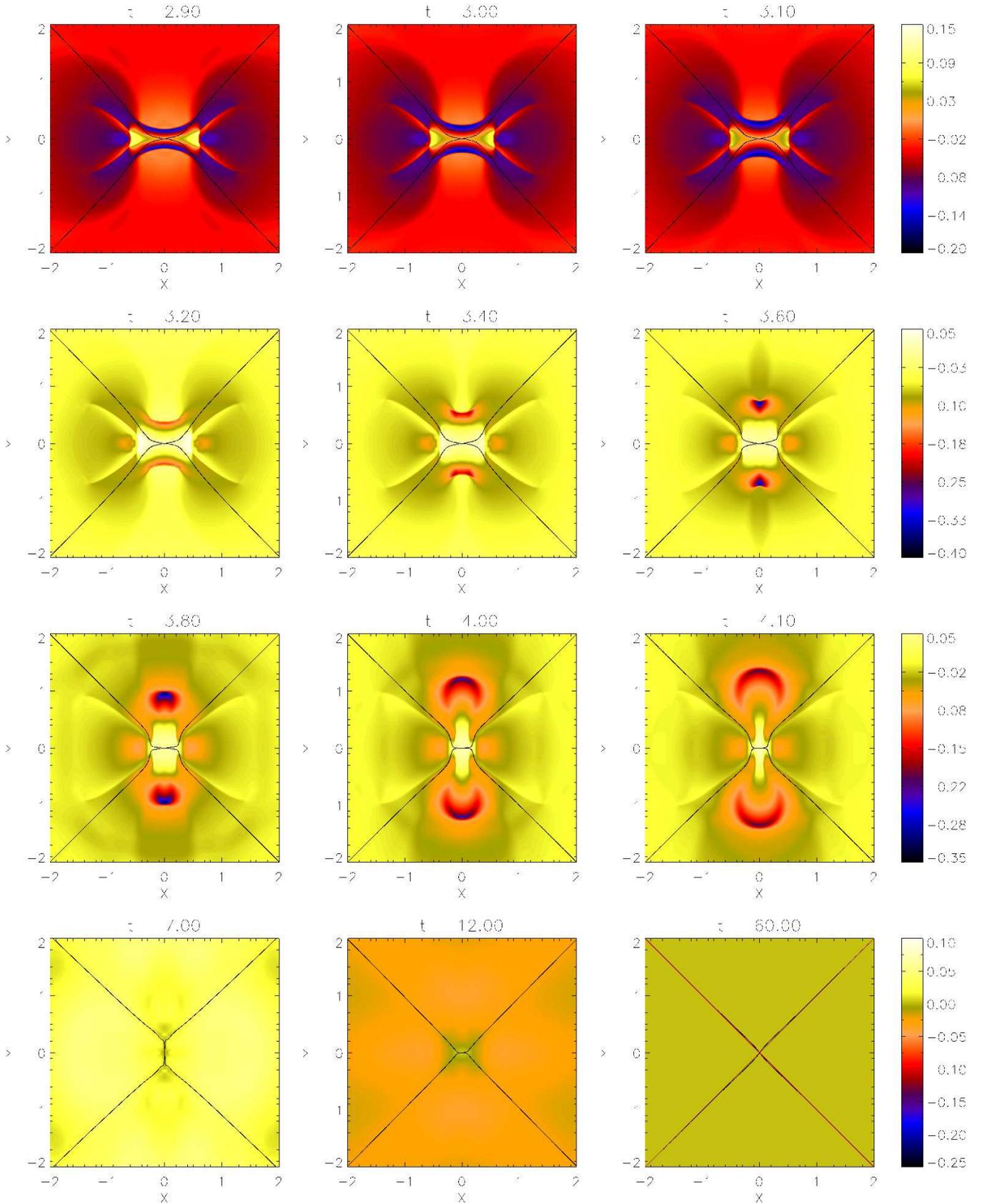}
\caption{Contours of ${\rm{v}}_\perp$ at various (Alfv\'en) times between $t=2.9$ and $t=60$. Note that the amplitude varies substantially throughout the evolution, and hence each row is assigned its own colour bar. The black lines denote the separatrices. The red lines at $t=60$ denote the potential separatrices.}
\label{figurenine}
\end{center}
\end{figure*}

The wave evolution between $1.4 \le t \le 2.8$ can be seen in Figure \ref{figurefive}. From the first row, $1.4 \le t \le 1.8$, we see that the asymmetry  seen in Figure \ref{figuretwo} has now lead to the formation of shock waves. From the second and third rows ($2 \le t \le 2.5$) of  Figure \ref{figurefive}, we see that the shocks above and below $y=0$ have started to overlap (starting  at $x\approx \pm 1$). This overlap leads to the development of hot jets.

\begin{figure}
\begin{center}
\includegraphics[width=3.75in]{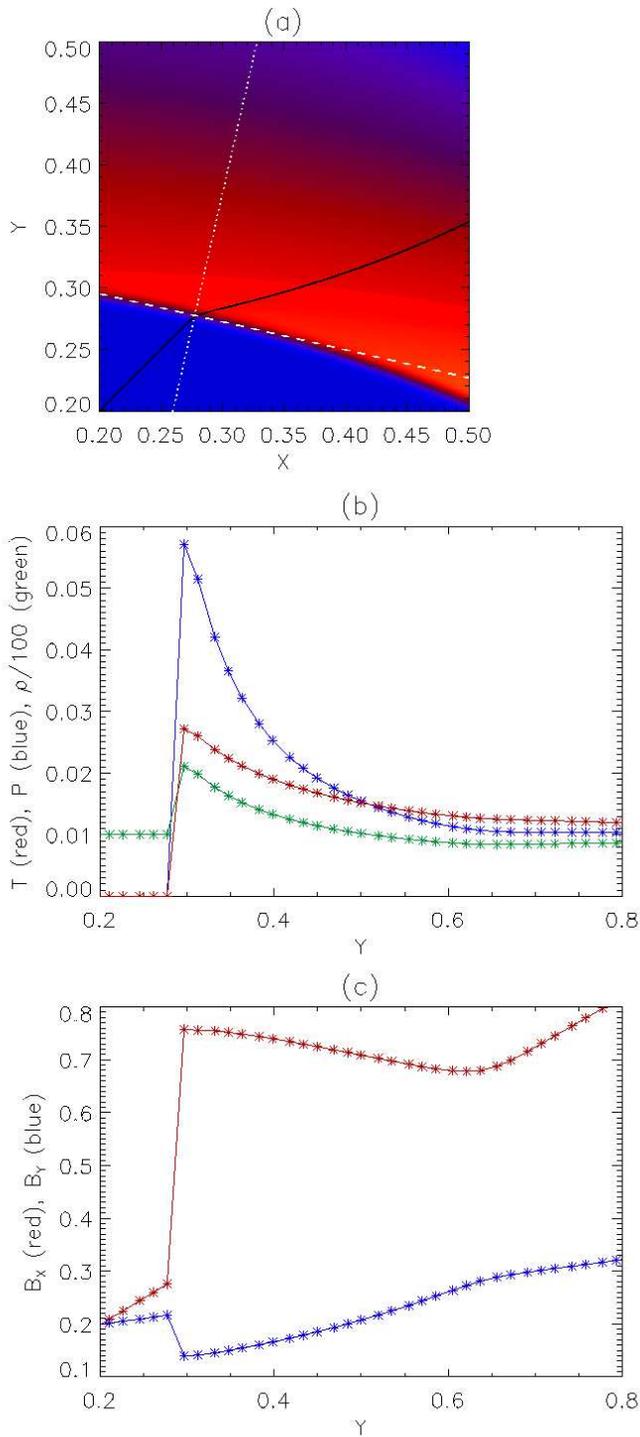}
\caption{({\emph{a}}) Contour of ${\rm{v}}_\perp$ at $t=2$ for $0.2 \le x,y \le 0.5$. The black line denotes the separatrix, the  dotted, $y= 4.382x-0.938$, and dashed, $y=-0.228x+0.341$, white lines denote the lines perpendicular and parallel to the shock front, respectively. ({\emph{b}}) Plots of Temperature, $T$ (red), Pressure, $P$ (blue), and density, $\rho / 100$ (green), and ({\emph{c}}) plots of $B_x$ (red) and $B_y$ (blue)  perpendicular to the shock front.}
\label{figuresix}
\end{center}
\end{figure}

The white box in Figure \ref{figurefive} at $t=2$ is analysed in Figure \ref{figuresix}a. By considering the physical quantities  along a line perpendicular to the shock front, we see that there is an abrupt increase in density, temperature and, consequently,  pressure (Figure \ref{figuresix}b). $B_x$ increases in magnitude, whereas $B_y$ decreases, and both preserve their original  directions (Figure \ref{figuresix}c). Hence, the shock makes ${\bf{B}}$ refract away from the normal and so we identify it as a fast oblique magnetic shock. In the idealised limit ${\bf{v}} \parallel {\bf{B}}$, this would be called a switch-on shock (see e.g. \S{1.5.2} in Priest \& Forbes \cite{magneticreconnection2000}). It is interesting to note that the shock has heated the plasma and so $\beta \neq 0$ in these locations.

\begin{figure}
\begin{center}
\includegraphics[width=3.75in]{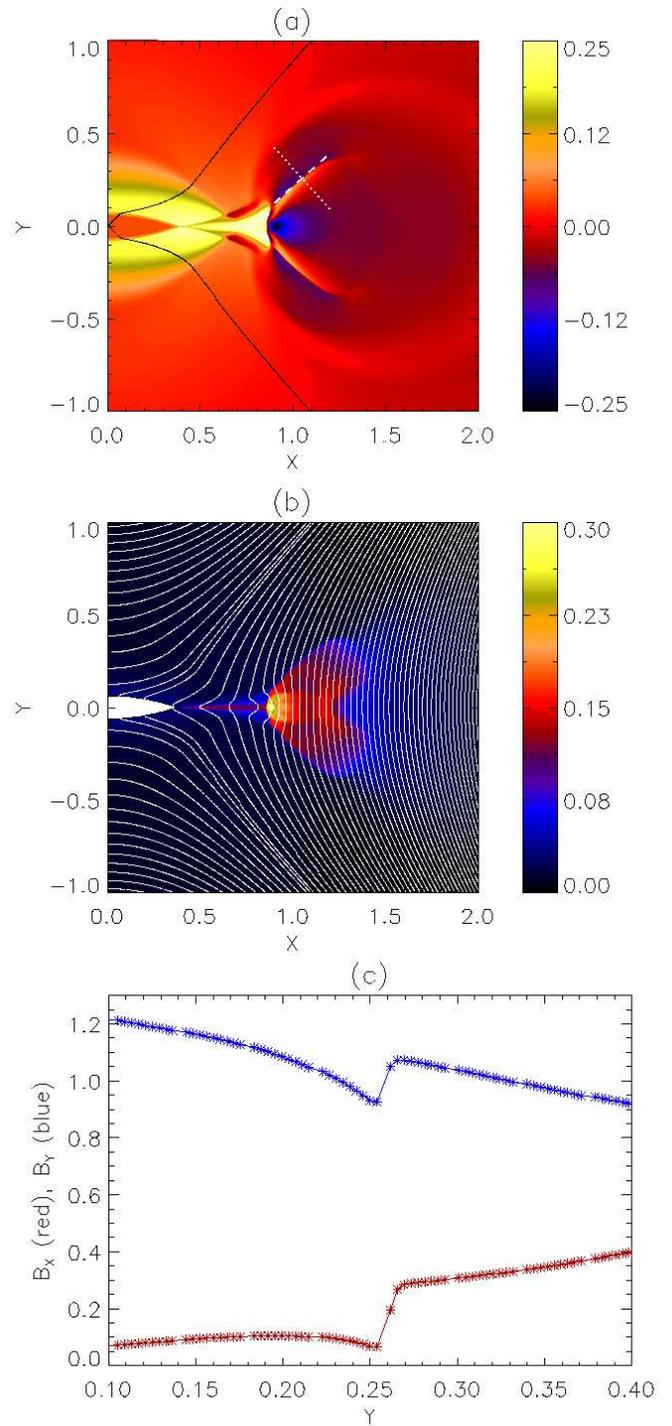}
\caption{({\emph{a}}) Contour of ${\rm{v}}_\perp$ at $t=2.4$ for $0\le x\le2 $, $-1\le y \le 1$. The black line denotes the separatrices, the  dotted, $y= -1.111x+1.427$, and dashed, $y=0.9x-0.685$, white lines denote the lines perpendicular and parallel to the shock front, respectively. ({\emph{b}}) Contour of temperature at $t=2.4$ for $0\le x\le2 $, $-1\le y \le 1$. The white lines denote magnetic fieldlines (thick white line  denotes the separatrices).  ({\emph{c}}) Plots of $B_x$ (red) and $B_y$ (blue) perpendicular to the shock front.}
\label{figureseven}
\end{center}
\end{figure}

The white box in Figure \ref{figurefive} at $t=2.4$ is analysed in Figure \ref{figureseven}. Figure \ref{figureseven}a shows a contour of ${\rm{v}}_\perp$ for $0\le x\le2 $, $-1\le y \le 1$, and we see that the overlap of the shock waves forms a triangular \lq{cusp}\rq{}, which we call the shock-cusp. Figure \ref{figureseven}b shows that these hot jets heat the plasma, substantially more than the previous shock heating (compare magnitudes of $T$ from Figure  \ref{figuresix}b and  Figure  \ref{figureseven}b). These hot jets also significantly bend the  magnetic fieldlines.

The hot jets set up new shock waves emanating from the shock-cusp, and by considering the changes of physical quantities perpendicular to the shock front (similar to before), we see that there is an abrupt increase in $B_x$ and decrease in $B_y$, although both preserve their original directions (Figure \ref{figureseven}c).  Thus, the shock makes ${\bf{B}}$ refract towards the normal, and we identify this as a slow oblique magnetic shock. In the idealised limit ${\bf{v}} \parallel {\bf{B}}$, this would be called a switch-off shock.

{{The structure of the hot jet is in good agreement with that described by Forbes (\cite{Terry}) (see his Figure 15). In addition to the slow shocks downstream of the tip of the jet, there is evidence of slow shocks along the sides of the jet upstream of the tip (see Figure \ref{figureseven}b). There are also kinks in the fieldlines at the tip of the jet, indicative of a fast oblique magnetic shock. Thus, the jet heating itself is accomplished by a combination of slow and fast shocks. The abrupt jump in temperature and magnetic field at the tip of the jet is also consistent with the jump predicted by Soward and Priest (\cite{Soward}).}}

The fourth row of  Figure \ref{figurefive} shows that the shocks, both the fast shocks and their overlap (the tails of the jets) have reached the null point (at $t=2.6$). The shocks have deformed the magnetic field such that the separatrices  now touch one another rather than intersecting at a non-zero angle (called \lq{cusp-like}\rq{} by Priest \& Cowley \cite{PC1975}). That the perturbation can reach the neutral point is a phenomenon not seen in {\cite{MH2004}} (where the perturbation reached the null after an infinite amount of time).

\begin{figure}
\begin{center}
\includegraphics[width=3.75in]{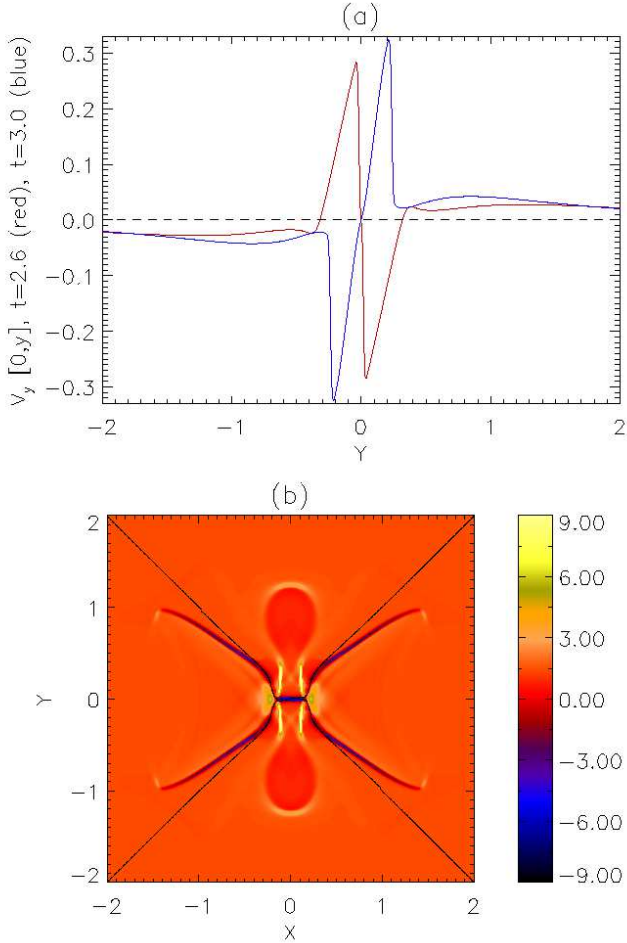}
\caption{({\emph{a}}) Plot of ${\rm{v}}_y(0,y)$ for $-2\le y \le 2$ at $t=2.6$ (red) and $t=3.0$ (blue). The perturbation has crossed the null point. The dashed line indicates ${\rm{v}}_y(0,y)=0$. ({\emph{b}})  Contour of $j_z$ at $t=4$.  The black lines denote the separatrices.}
\label{figureeight}
\end{center}
\end{figure}

After time $t=2.6$, we see that the shock wave can now pass through the null point: entirely different behaviour to that seen in {\cite{MH2004}}. This is more clearly seen in Figure \ref{figureeight}a, which shows a plot of ${\rm{v}}_y (0,y)$ at $t=2.6$ (just before null point is crossed) and $t=3.0$ (after null has been crossed).

\newpage

\clearpage

\newpage

\clearpage

\newpage

\clearpage
\newpage

\newpage

\subsection{Oscillatory Behaviour ${\rm{v}}_\perp$ $(4.1\le t\le60)$}\label{section3.3}

\begin{figure}[t]
\begin{center}
\includegraphics[width=3.75in]{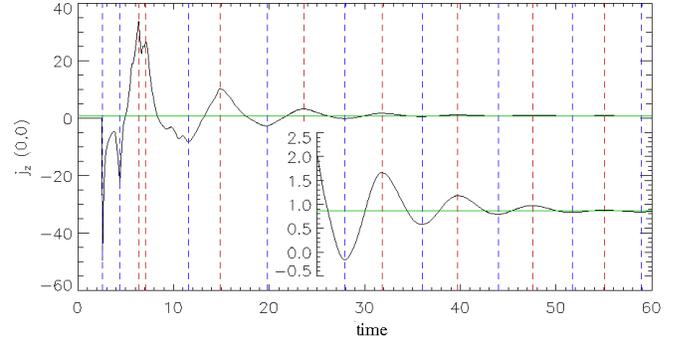}
\caption{Plot of time evolution of $j_z(0,0)$ for $0\le t \le 60$. Insert shows plot of time evolution of $j_z(0,0)$ for $25\le t \le 60$ (same $x-$axis, different $y-$axis).  The dashed lines indicate maxima (red) and minima (blue) in the system and the green line shows $j_z(0,0)=0.8615$.}
\label{figureten}
\end{center}
\end{figure}

The evolution of  ${\rm{v}}_\perp$ for $2.9 \le t \le 60$ can be seen in Figure  \ref{figurenine}.  We can see that the evolution proceeds  in two separate ways. Firstly, some of the wave now escapes the system: this propagation can be seen above and below the location of the null point. Secondly, the  (deformed) neutral point itself continues to collapse and forms a horizontal current sheet. This can be clearly seen at $t=4$, and contours of $j_z$ are shown in  Figure \ref{figureeight}b. Here, current density can be seen at the four slow oblique magnetic shocks, along (approximately) $-0.2\le x \le 0.2, y=0$ (i.e. a horizontal current sheet), at the locations of our two shock-cusps  and due to the wave propagating away. Thus, we see that current density forms at several locations. The formation of a current sheet was not seen in {\cite{MH2004}} (which reported the formation of a current density line at the null point).

 Returning to Figure \ref{figurenine}, we see that this horizontal current sheet shortens in length ($t=4.1$). This is because the jets to the left and right of the origin heat the plasma, which begins to expand. This expansion  squashes the horizontal current sheet, forcing the separatrices apart. The (squashed) horizontal current sheet then returns to a  \lq{cusp-like}\rq{} null point which, due to the continuing expansion from the heated plasma, in turn forms a vertical current sheet ($t=7$). The evolution now proceeds through a series of horizontal and vertical current sheets and displays oscillatory behaviour (as seen by Craig \& McClymont \cite{CraigMcClymont1991}). At $t=60$, we see that the majority of the initial wave pulse has propagated away from the X-point, and that the associated velocities are negligible. It is also interesting to note that the final state ($t=60$) is non-potential; the separatrices for $t=0$ are overplotted in red and there is clearly a small offset. This is due to the finite amount of current left in our system (see below).

The oscillatory nature of the system can be clearly seen from the time evolution of  $j_z(0,0)$ shown in Figure \ref{figureten}. The red / blue lines indicate  maxima / minima in the system and the green  line shows $j_z(0,0)=0.8615$, which is the limiting value of the oscillation. We can see that there is no current density at $x=y=0$ (i.e. the location of the potential null point) before $t=2.6$. This is as expected since our initial wave pulse does not reach the null before this time. At $t=2.6$ we see a strong spike in the current density: this is the result of our nonlinear wave reaching the null point for the first time (see Figure \ref{figurefive} at $t=2.6$). At $t=4.4$, the current density associated with the triangular shock-cusps (from the slow oblique magnetic shocks) reaches the origin; giving rise to a second sudden spike in current density.

At $t=6.4$, there is a third large spike in current density (opposite sign to previous two peaks). This indicates the formation of the first vertical current sheet. The peak shortly following this (at $t=7.4$) is due to the current density associated with the triangular shock-cusps from this vertical current sheet reaching the origin.

The next peak in current density occurs at $t=11.8$. This represents the formation of a second horizontal current sheet (see  Figure \ref{figurenine} at $t=12$). However, there is no secondary peak associated with this horizontal current sheet, unlike the previous two current sheets.

From Figure  \ref{figureten} we can see that at later times, the cycle of maxima and minima continues, with each maxima (red lines) associated with the formation of vertical  current sheet and each minima (blue) associated with a horizontal one. This oscillation has a decreasing period, and the  time taken to go from a horizontal to vertical current sheet is shorter than vice-versa. This is because the system finds it easier to form vertical current sheets due to the asymmetric heating around the null point (see below).

We can also see that the oscillation in current density is tending to a constant value of $j_z(0,0)=0.8615$. This is in agreement with our statement above that the final state is non-potential. This is because at the end of the simulation, the plasma to the left and right of the neutral point is hotter than above and below, due to the hot jets that formed and heated the (initially cold) plasma after $t=2$. Consequently,  the plasma pressure in greater to the left and right of the null and hence the system finds it easier to form vertical current sheets (due to the asymmetric heating). The final state (Figure  \ref{figurenine} at $t=60$) shows that the X-point is very slightly closed up in the vertical direction, in agreement with $j_z$ tending to a small positive value.

\begin{figure}
\begin{center}
\includegraphics[width=3.75in]{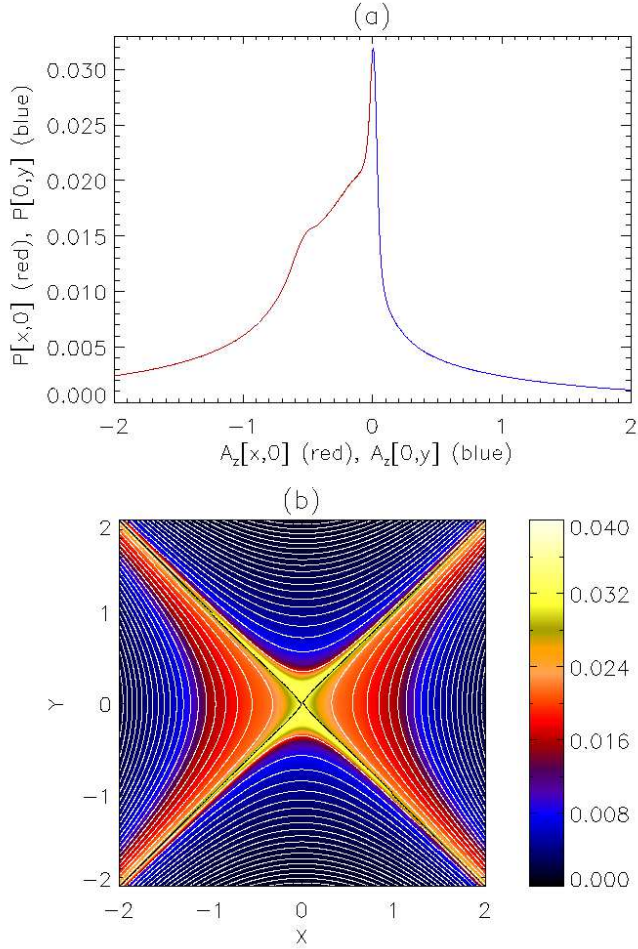}
\caption{$(a)$ Plot of $P[x,0]$ (red) against $A_z[x,0]$ and plot of  $P[0,y]$ (blue) against  $A_z[0,y]$. $(b)$ Contour of plasma pressure at $t=60$. Contours of $A_z(t=60)$ are overplotted in white and the separatrices are in black.}
\label{figuretenBBB}
\end{center}
\end{figure}

If the final state is in force-balance, we would expect the pressure to be a function of $A_z$, i.e. the plasma pressure should lie along contours of $A_z$ if $P=P(A_z)$. Figure \ref{figuretenBBB}a shows that a plot of $A_z(x,0)$ and $A_z(0,y)$ against $P(x,0)$ (red) and  $P(x,0)$ (blue) yields a single curve, and from Figure \ref{figuretenBBB}b, we can see that the agreement between the pressure and contours of $A_z(t=60)$ is excellent.  This implies that although the final state is non-potential, it is still in force-balance.

\newpage

\section{Reconnection}\label{section4}

\begin{figure*}
\begin{center}
\includegraphics[width=7.5in]{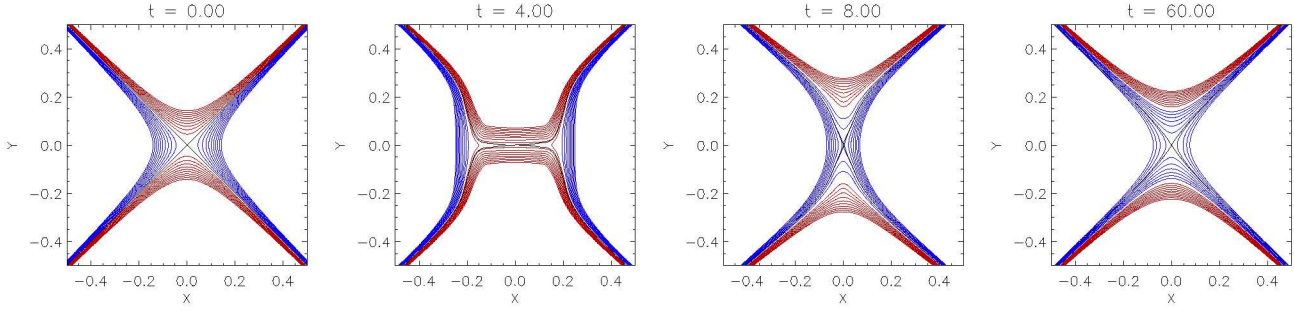}
\caption{Selection of fieldlines in our system at  times $t=0,4,8 \;\&\; 60$. Red fieldlines originate from  $x \in [-20, -19.99], y=\pm20$ and blue fieldlines originate from  $x=\pm20, y \in [-20, 19.99]$, all  with equal separation of $0.01$. The separatrices is plotted in black. The same fieldines are shown in each subfigure.}
\label{figuretwelve}
\end{center}
\end{figure*}

\begin{figure*}
\begin{center}
\includegraphics[width=7.5in]{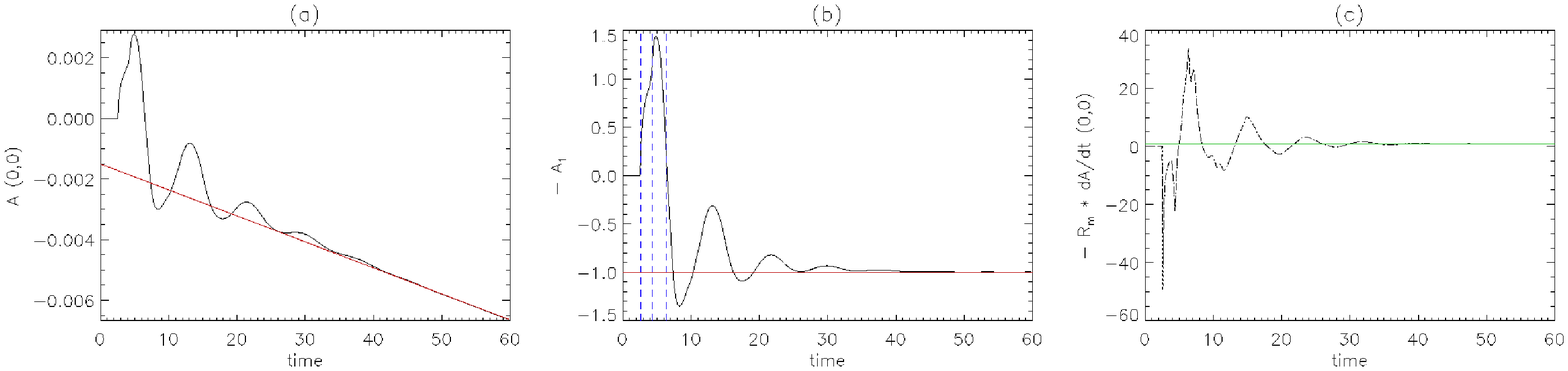}
\caption{$(a)$ Plot of the time evolution of $A_z(0,0)$, where we have overplotted a straight line fit: $A_z=-10^{-5}\left(8.615t+148.434\right)$. $(b)$ Plot of the rate of change of $A_z$ at $x=y=0$ with straight line trend removed: $-A_1 = 10^5 A_z(0,0) / \left(8.615t+148.434\right)$, where $t=2.6,4.4\: \& \:6.4$ are shown in blue. $(c)$ (Dashed line) Plot of the time evolution of the reconnection rate, i.e. $-R_m \left. {\partial A_z} / {\partial t}\right| {(0,0)}$, (dotted line) overplot of evolution of $j_z(0,0)$. {{The green line shows  $-R_m \left. {\partial A_z} / {\partial t}\right| {(0,0)}=0.8615$. }}  }
\label{figurethirteen}
\end{center}
\end{figure*}

From Section \ref{section3.3}, it is clear that we have oscillatory behaviour in our system. However, do we have reconnection?  We demonstrate that reconnection is occurring via the following two pieces of evidence: firstly, in Figure \ref{figuretwelve} we have plotted a selection of fieldlines in our system at four different time slices, where red fieldlines originate from  $x \in [-20,-19.99], y=\pm20$ and blue fieldlines originate from  $x=\pm20, y \in [-20,-19.99]$, all with equal separation of $0.01$. The separatrices (which change in time) are plotted in black. In order to clearly show the results, we have plotted a subsection of our computational box: $x,y \in [-0.5,0.5]$.

At $t=0$, all the red fieldlines from the top / bottom left corner connect to the top / bottom right corner, and all the blue fieldlines connect from the top  left / right corner to the  bottom left / right  corner. The (potential) separatrices pass through $x, y=[\pm20,\pm20]$ and the origin, as expected from equation (\ref{Xpoint}), and separate the red and blue fieldlines. Thus, if we have a change in connectivity, we should see a red / blue fieldline cross the (evolving) separatrices. Since we have reflecting boundaries $({\bf{v=0}})$ and the value of the vector potential, $A_z$, on the boundaries does not change with time, we can be confident that we are always plotting  the same fieldlines in each subfigure.

Figure \ref{figuretwelve} ($t=4$) shows our choice of magnetic fieldlines shortly after the formation of the first horizontal current sheet (compare to Figure \ref{figurefive} at $t=4$). Here, we can see that one of our red fieldlines has crossed the separatrices; indicating a change in connectivity. Figure \ref{figuretwelve} ($t=8$) shows our choice of magnetic fieldlines shortly after the formation of the first vertical  current sheet, and now we see that some of the blue lines have crossed the separatrices and the red fieldlines are once again all on the same side of the separatrices. At the end of our simulation ($t=60$), we see that more blue fieldlines have crossed the separatrices. Thus, throughout the evolution of our system, we have several changes in connectivity, giving us qualitative evidence for reconnection.

Quantitative evidence for reconnection in our system can be seen in Figure \ref{figurethirteen}. Here, we see a plot of the time evolution of $A_z(0,0)$, where changes in the vector potential at the origin indicate changes in connectivity.  Figure \ref{figurethirteen}a shows a plot of the time evolution of $A_z(0,0)$. Before $t=2.6$,  $A_z(0,0)=0$ as expected (no disturbance to the null point). After this time, we see that  $A_z(0,0)$ oscillates but also displays a clear trend that is tending towards a straight line: $A_z=-10^{-5}\left(8.615t+148.434\right)$. This straight line is associated with the final current density remaining in our system, i.e.:
\begin{eqnarray}
\frac{\partial A_z}{\partial t} = \frac{1}{R_m} \nabla ^2 A_z  =  - \frac{1}{R_m} j_z \;. \label{Az}
\end{eqnarray}
Hence, if our final state contains constant current density, say $j_{C=1}$, then we expect $A_z$ to change linearly in time. Assuming $j_z\rightarrow j_{C=1}$ and integrating equation (\ref{Az}) and comparing this to our straight line, we see that $j_{C=1} = 0.8615$, which is in excellent agreement with our estimate from Figure \ref{figureten}.

Figure \ref{figurethirteen}b shows a plot of the time evolution of $A_1 = -10^5 A_z(0,0) / \left(8.615t+148.434\right)$, i.e. we have removed the straight line trend. We have actually plotted  $-A_1$ to aid comparison between Figures \ref{figurethirteen}a and \ref{figurethirteen}b. We can clearly see the oscillatory nature of $A_1(0,0)$ and that it tends towards a constant value of $-1$ (which gives confidence that our straight line fit is appropriate). We can also see that $A_1(0,0)$ undergoes significant changes at $t=2.6,4.4\; \&\; 6.4$ (overplotted in blue), where these times correspond to the nonlinear wave reaching the X-point, the first triangular shock-cusps reaching the null and the formation of the first vertical current sheet, respectively.

Figure \ref{figurethirteen}b shows us that $A_1(0,0)$ is continuously changing as it tends to $-1$, and thus reconnection is continuously occurring until this time ($t\approx 50$).  Thus, whenever $-A_1(0,0)$ oscillates through $-1$, we are increasing or decreasing flux on one side or the other of our final state.

Finally, we can calculate the reconnection rate in our system, defined as $\left.{\partial A_z} / {\partial t}\right|_{(0,0)}$, and we have  plotted  $ - R_m \left.{\partial A_z} / {\partial t}\right|_{(0,0)}$    in  Figure \ref{figurethirteen}c. We can clearly see the reconnection rate varies throughout the simulation and tends towards a constant value. However, from equation (\ref{Az}) we see that the reconnection rate is the same as  $- j_z / R_m$ and thus is expected to tend towards  $- j_{C=1} / R_m$. Indeed we have overplotted the current density evolution (dotted line) and the agreement is excellent.

\newpage

\section{Amplitudes C=0.5 and C=2}\label{section5}

\begin{figure*}
\begin{center}
\includegraphics[width=7.5in]{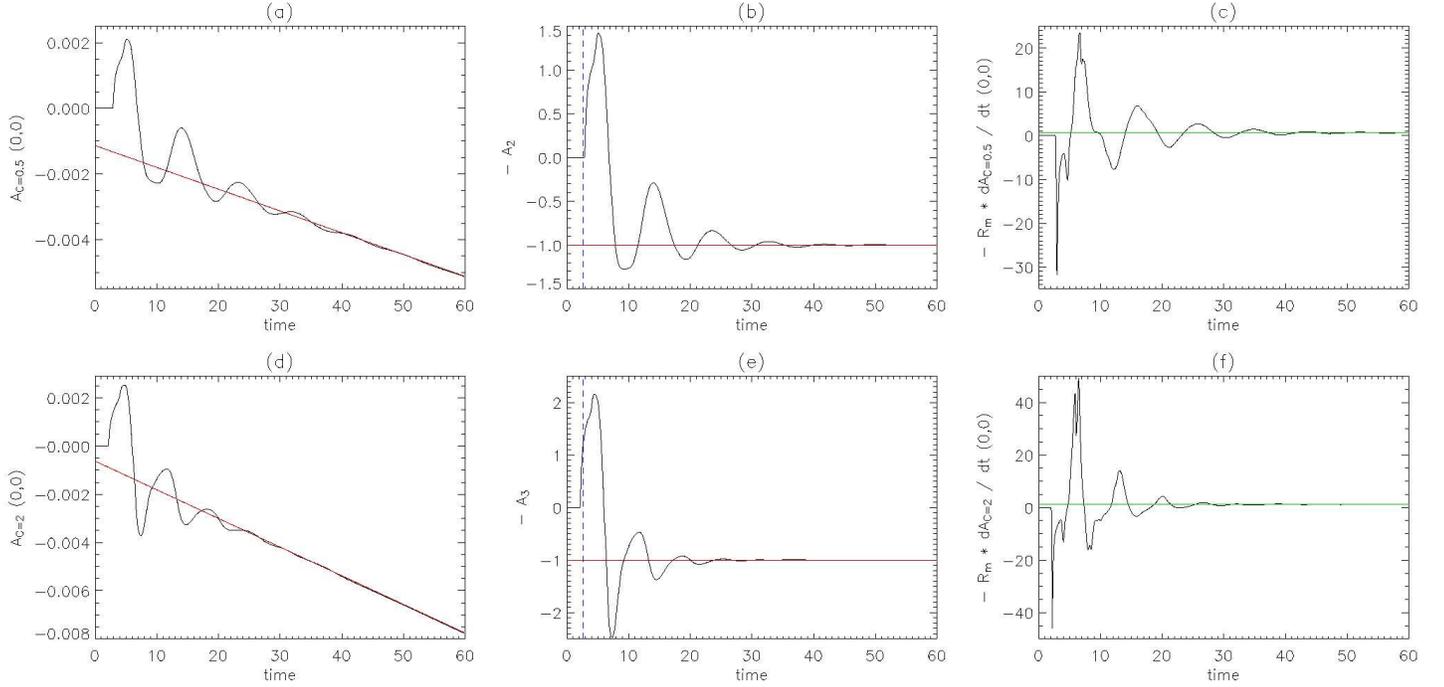}
\caption{$(a)$ Plot of the time evolution of $A_{C=0.5}(0,0)$, where we have overplotted a straight line fit: $A_{C=0.5}=-10^{-5}\left( 6.672t +113.452\right)$. $(b)$ Plot of the rate of change of detrended $A_{C=0.5}$ at $x=y=0$, where $-A_2 = 10^5 A_{C=0.5}(0,0) / \left( 6.672t +113.452\right)$, where $t=2.6$ is shown in blue. $(c)$ Plot of the time evolution of the reconnection rate, i.e. $-R_m \left. {\partial A_{C=0.5}} / {\partial t}\right| {(0,0)}$, {{where the green line indicates  $-R_m \left. {\partial A_{C=0.5}} / {\partial t}\right| {(0,0)}= 0.6672$}}. $(d)$ Plot of the time evolution of $A_{C=2}(0,0)$, where we have overplotted a straight line fit: $A_{C=2}=-10^{-4}\left(1.194t+6.242\right)$. $(e)$ Plot of the rate of change of detrended $A_{C=2}$ at $x=y=0$, where $-A_3 = 10^4 A_{C=2}(0,0) / \left(1.194t+6.242\right)$, where $t=2.6$ is shown in blue. $(f)$ Plot of the time evolution of the reconnection rate, i.e. $-R_m \left. {\partial A_{C=2}} / {\partial t}\right| {(0,0)}$, {{where the green line indicates $-R_m \left. {\partial A_{C=2}} / {\partial t}\right| {(0,0)}= 1.1943$.    }}  }
\label{figurefourteen}
\end{center}
\end{figure*}

In this section, we examine the effect of changing  the amplitude, $C$, of our initial condition in ${\rm{v}}_\perp$ (equation \ref{ICs}). Here, we consider $C=0.5$ and $C=2$, and the results can be seen in Figure \ref{figurefourteen}. Comparing to Figure  \ref{figurethirteen}, we can see that the overall behaviour is very similar that of $C=1$, where  we observe oscillatory reconnection and a cycle of horizontal and vertical current sheets. Figure \ref{figurefourteen}a shows a plot of the time evolution of  $A_{C=0.5}(0,0)$, i.e. the flux function at the origin for $C=0.5$. We see that  $A_{C=0.5}(0,0)$ oscillates but displays a clear downwards trend that is tending to a straight line: $A_z=-10^{-5}\left( 6.672t +113.452\right)$. As explained in Section \ref{section4}, this straight line is associated with the final current density remaining in our system. Figure \ref{figurefourteen}b shows a plot of the time evolution of $-A_2 =  10^5 A_{C=0.5}(0,0) /  \left( 6.672t +113.452\right)$, i.e. we have detrended $A_{C=0.5}(0,0)$. We can clearly see the  oscillatory nature of $A_{C=0.5}(0,0)$. As before, the repeated  changes in $A_{C=0.5}(0,0)$ implies that reconnection is continuously occurring in our system.

We can also calculate the reconnection rate in the $C=0.5$ system, defined as $\left.{\partial A_{C=0.5}} / {\partial t}\right| {(0,0)}$ and we have plotted $-R_m \left. {\partial A_{C=0.5}} / {\partial t}\right| {(0,0)}$ in Figure \ref{figurefourteen}c. We see that the reconnection rate varies throughout the simulation and tends to a constant value. From equation (\ref{Az}), we see that the reconnection rate is also an excellent measure of the current density in the system, tending to a final current density of $j_{C=0.5}= 0.6672$. Comparing to Figure \ref{figurethirteen}c, we see that the current evolution is comparable in nature but slightly  smaller in magnitude. In addition,  the final current left in the system is smaller than  $j_{C=1}=0.8615$.

 Figure \ref{figurefourteen}d shows a plot of the time evolution of  $A_{C=2}(0,0)$, i.e. the flux function at the origin for $C=2$. As before, we see that  $A_{C=2}(0,0)$ oscillates and  displays a clear downwards trend that is tending to a straight line: $A_z=  -10^{-4}\left(1.194t+6.242\right)$.  Figure \ref{figurefourteen}e shows a plot of the time evolution of  $-A_3 = 10^4 A_{C=2}(0,0) / \left(1.194t+6.242\right)$. Again, we can clearly see the oscillatory nature of the system. Figure \ref{figurefourteen}d shows the reconnection rate, and therefore the current density evolution in the $C=2$ system. Comparing to Figure \ref{figurethirteen}c, we see  the current evolution is comparable in nature but slightly larger in magnitude. The final current left in the system,  $j_{C=2}=1.1943$, is larger than  $j_{C=1}=0.8615$.

We have also indicated $t=2.6$ in Figures \ref{figurefourteen}b and\ref{figurefourteen}e, i.e. the time taken for the nonlinear wave to first reach the X-point in the $C=1$ system. We see that in the $C=0.5$ / $C=2$ system, the wave reaches the X-point after / before this time, since the shock forms earlier and travels faster the larger the value of $C$.

\newpage

\section{Conclusions}\label{section:conclusions}

This paper describes an investigation into the nature of nonlinear fast magnetoacoustic waves in the neighbourhood of a 2D magnetic X-point. We have solved the compressible and resistive MHD equations using a Lagrangian remap, shock capturing code ({\emph{LARE2D}}). We consider a circular, sinusoidal pulse in ${\rm{v}}_\perp$ as our initial condition in velocity (equation \ref{ICs}), which naturally splits into two waves and we focus on the wave travelling towards the null point. Implemented damping regions remove the kinetic energy from the outgoing wave. Initially, we consider an incoming wave of amplitude $C=1$.

Between $0\le t \le 1$, we find that the incoming wave propagates across the magnetic fieldlines and keeps its initial pulse profile (an annulus). The annulus contracts as the wave approaches the null point, and this is the same refraction behaviour reported in \cite{MH2004}, due to the spatial variation of the (equilibrium) Alfv\'en speed.  We also note that the incoming wave pulse is developing an asymmetry: the wave peaks are propagating faster (relative to the footpoints) in the $y-$direction than in the $x-$direction. The incoming wave develops discontinuities, where in the $y-$ / $x-$ direction the wave peak / trailing footpoint is catching up with the leading footpoint / wave peak. The development of shocks around $t=1$ was not reported in \cite{MH2004}, as it is a nonlinear effect, and arises because of our choice of a velocity initial condition: in the nonlinear regime, specifying an initial condition in velocity also prescribes a background velocity profile (Figure \ref{figurefour}). This phenomenon is demonstrated for a simple system in Appendix \ref{AppendixB}. In addition, our initial condition in ${\rm{v}}_\perp$ appears to excite the $m=0$ mode, but this corresponds to the $m=2$ mode in cartesian components.

For $1.4 \le t \le 2.8$, we follow the asymmetry reported above in the development  of fast oblique magnetic shock waves. By considering the physical quantities along a line perpendicular to the shock front, we found an abrupt increase in density, temperature and (consequently) pressure. It is interesting to note that the shock has heated the initially $\beta=0$  plasma, thus creating $\beta \neq 0$ at these locations.

At $t\approx 2$, we observed that the shocks above and below $y=0$ began to overlap (starting at $x\approx \pm1$), forming a triangular \lq{cusp}\rq{}: the shock-cusp. This led to the development of hot jets, which substantially heated the local plasma and significantly bent the local magnetic fieldlines. The hot jets set up slow oblique magnetic  shock waves emanating from the shock-cusp. {{In addition, there is evidence of slow shocks along the sides of the jet upstream of the tip and we see kinks in the fieldlines at the tip of the jet, indicative of a fast shock.  Thus, the jet heating itself is accomplished by a combination of slow and fast shocks. It is interesting to note that the jet has a bimodal structure consisting of a hot, narrow jet incased within a broader, lower temperature jet, which is a feature that is not predicted by steady-state reconnection theory.  }}

The nonlinear wave, both the fast shocks and their overlap (the tails of the jets), reach the null point at $t=2.6$. The shocks have deformed the magnetic field such that the separtrices now touch one another rather than intersecting at a non-zero angle (called \lq{cusp-like}\rq{} by Priest \& Cowley \cite{PC1975}). {{However, as can be seen at later times, the separatrices continue to evolve and so this osculating field structure is not sustained for any length of time.}} That the perturbation can reach the null point is a phenomenon not seen in \cite{MH2004}. The nonlinear wave then passes through the null point, again entirely different behaviour to that seen in \cite{MH2004}.

After $t=2.6$, The evolution subsequently  proceeds in two separate ways. Firstly, some of the wave now escapes the system, corresponding to the wave that has passed through the null point. Secondly, the (deformed) null point itself continues to collapse and forms a horizontal current sheet. Current density exists in this horizontal current sheet, and also  at the four slow oblique magnetic shocks, at the location of the shock-cusps and in the wave propagating away (as opposed to in the linear system, where the current density accumulated at the null point).

After $t=4$, the evolution proceeds as follows: the jets to the left and right of the origin continue to heat the plasma, which in turn expands. This expansion squashes and shortens the horizontal current sheet, forcing the separatrices apart. The (squashed) horizontal current sheet then returns to a  \lq{cusp-like}\rq{} null point which, due to the continuing expansion from the heated plasma, in turn forms a vertical current sheet. The evolution then proceeds through a series of horizontal and vertical current sheets and displays oscillatory behaviour (as reported by Craig \& McClymont \cite{CraigMcClymont1991}).

At the end of our simulation ($t=60$), we see that the majority of the wave pulse has propagated away from the null point, and that the associated velocities are negligible. It is also interesting to note that our final state is non-potential: there is a finite amount of current density left in our system ($j_{C=1}=0.8615$). This non-potential state occurs because the plasma to the left and right of the null point is hotter than that above and below, due to the hot jets that formed and heated the initially $\beta=0$ plasma after $t=2$. Consequently, the plasma pressure is greater to the left and right and hence the final state shows that the X-point is slightly closed up in the vertical direction. This also explains why the time taken to go from a horizontal to vertical current sheet is shorter than the reverse: the system finds it easier to form vertical current sheets due to this asymmetric heating around the null. We also found that the pressure is approximately a function of the vector potential, i.e. $P=P(A_z)$, in the final state, indicating that although the final state is non-potential, it is still approximately in force balance. Of course, the system will eventually return to a potential state due to diffusion, but this will occur on a far greater timescale than that of our simulation ($t_{\rm{diffusion}} \sim R_m=10^4$).

We provide two pieces of evidence for reconnection in our system. Qualitatively, we observed changes in fieldline connectivity (Figure \ref{figuretwelve}) and quantitatively we looked at the evolution of the vector potential at the origin. We found that $A_z(0,0)$ oscillates and displays a clear trend that was tending towards a straight line: $A_z=-10^{-5}\left(8.615t+148.434\right)$. This straight line is associated with the final current density left in the system. Thus, since we have both oscillatory behaviour in our system {\emph{and}} evidence for reconnection, we conclude that the system displays {\emph{oscillatory reconnection}} (detailed by Craig \& McClymont \cite{CraigMcClymont1991}, and reported  more recently by Murray et al. \cite{Murray}).

We then extended our study to look at the effect of changing the amplitude, $C$, of our initial condition (equation \ref{ICs}). Considering both $C=0.5$ and $C=2$, we found that the overall behaviour was very similar to the $C=1$ study, i.e. oscillatory reconnection and a cycle of horizontal and vertical current sheets. Looking at the evolution of $A_z(0,0)$ for the two systems, labelled $A_{C=0.5}$ and $A_{C=2}$, we saw that both oscillated and displayed a clear downward tend. Again, this was associated with the final current left in the system, where $j_{C=0.5}=0.6672$ was smaller than $j_{C=1}=0.8615$, which were both less than $j_{C=2}=1.1943$. Thus, we conclude that a larger initial amplitude results in a larger amount of current being left in the system at the end of the simulation, i.e. in the non-potential final state. Note that in the linear limit, $j_{C=0.001}=0$, since the wave cannot reach the null point in a finite time.

Thus, we can now fully answer our three original questions (posed in \S\ref{section1}):
\begin{itemize}
\item[(1)]{{\emph{Does the fast wave now steepen to form shocks, and can these propagate across or escape the null?}}}\par{The nonlinear behaviour is completely different to the linear regime. We observe the formation of both fast and slow oblique magnetic shocks, and the nonlinear wave can now cross, and thus escape, the null point. We have also seen that the shocks (asymmetrically) heat the plasma such that $\beta\neq 0$.}
\item[(2)]{{\emph{Can the refraction effect drag enough magnetic field into the null to initiate X-point collapse or reconnection?}}}\par{The nonlinear wave deforms the X-point into a \lq{cusp-like}\rq{} point, which in turn collapses into a horizontal current sheet. Expanding plasma to the left and right squashes the horizontal current sheet and the system evolves through a series of horizontal and vertical current sheets. Changes in the value of the vector potential at the origin as well as changes in connectivity demonstrate we have reconnection in our system. Our final state is non-potential, but in  force balance. Larger amplitudes in our initial condition correspond to larger values of the final current left in the system.}
\item[(3)]{{\emph{Has the rate of current density accumulation changed, and is the null still the preferential location of wave heating?}}}\par{The current density now accumulates at many locations, such as along horizontal or vertical current sheets, along slow oblique magnetic shocks and at the location of shock-cusps. Current density can now also leave the system since the nonlinear wave can escape the null point. This was not the case in the linear regime, where all the current density accumulated at the null point exponentially in time.}
\end{itemize}

The aim of this  paper is  to contribute to the understanding of how nonlinear MHD waves behave in inhomogeneous, magnetised environments. Future work in this area will consider the consequences of different initial conditions, such as utilising a localised increase in pressure or internal energy in a $\beta\neq 0$ plasma. Such an initial condition is expected to introduce slow waves and create a region around the null point where the sound speed and Alfv\'en speed become comparable in magnitude, and hence a layer where mode conversion could occur (e.g. Cally \cite{Cally}; McLaughlin \& Hood \cite{MH2006b}; McDougall \& Hood \cite{Dee}). Finally, this work will also be extended to nonlinear wave behaviour in the neighbourhood of a 3D null point, and compared to linear work by, for example,  Galsgaard et al. (\cite{Klaus}), Pontin \& {Galsgaard} (\cite{P1}), Pontin et al. (\cite{P2}) and McLaughlin et al. (\cite{MFH2008}).

\begin{acknowledgements}
{{The authors would like to thank the referee, Terry Forbes, for his valuable comments that helped to improve this paper}}. JAM and IDM acknowledge financial assistance from the Leverhulme Trust and Royal Society, respectively. JAM also wishes to thank Michelle Murray, Valery Nakariakov and Neil Symington for helpful and insightful discussions. The computational work for this paper was carried out on the joint STFC and SFC (SRIF) funded cluster at the University of St Andrews (Scotland, UK).
\end{acknowledgements}

\newpage

%


\appendix

\section{Linear Regime}\label{AppendixA}

\begin{figure*}
\begin{center}
\includegraphics[width=7.5in]{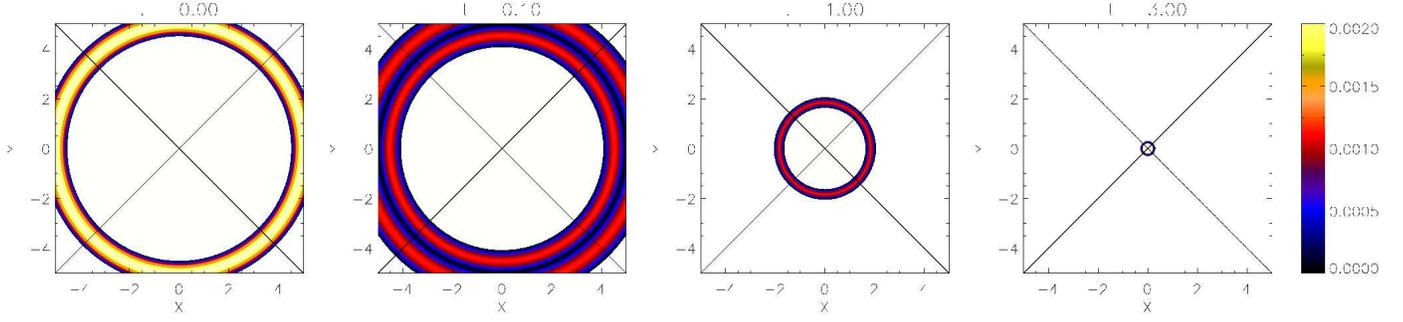}
\caption{Contours of ${\rm{v}}_\perp$ for a fast wave pulse initially located at a radius $r=5$, and its resultant propagation at (Alfv\'en) times $t=0, 0.1,1$ \& $3$. At $t=0.1$, we can see the initial pulse has split into two, oppositely travelling wave pulses. The black lines denote the separatrices and the null point is located at the origin.}
\label{figureappendixA}
\end{center}
\end{figure*}

In this appendix, we numerically solve equations (\ref{MHDequations}) subject to initial condition (\ref{ICs}) and set $C=0.001$. This recovers the linear results for the fast MA wave seen in {\cite{MH2004}}. This appendix has been included because ($i$) it will be useful to directly compare the linear and nonlinear systems subject to the same initial condition, and ($ii$) {\cite{MH2004}} described a wave pulse driven in from one of the boundaries,which is somewhat different from the setup studied in the present paper. The results described below (simulations using the {\emph{LARE2D}} numerical code) are identical to those in {\cite{MH2004}} (simulations using  a two-step Lax-Wendroff numerical scheme), and the results can be seen in Figure \ref{figureappendixA}. We find that the initial condition, as expected, splits into an outgoing wave and an incoming wave. We concentrate on the incoming wave. This wave propagates across the magnetic fieldlines and travels at the Alfv\'en speed, and we  identify it as a linear fast MA wave, in a cold plasma ($\beta=0$).  The linear fast wave keeps its initial pulse profile, i.e. an annulus, and the maximum amplitude remains at $C$, and the annulus contracts as the wave approaches the null point. Since the Alfv\'en speed is spatially varying, a refraction effect focuses the wave into the null point.

As the length scales decrease, there is a build up of current density, growing exponentially in time, while the velocity remains finite in magnitude.  Ohmic heating occurs preferentially at the null point. The refraction effect  and the preferential heating at the null point are the key results for linear, $\beta=0$ plasma fast wave propagation ({\cite{MH2004}}). Note that since the magnetic field, and hence the Alfv\'en speed, is zero at the null point, the fast wave cannot cross the null point and never actually reaches it.

\section{Isothermal 1D HD equations}\label{AppendixB}

\begin{figure*}
\begin{center}
\includegraphics[width=7.5in]{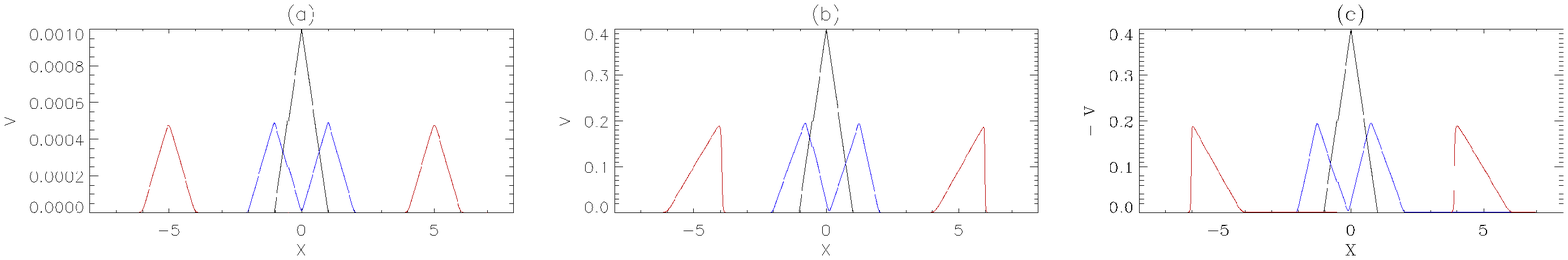}
\caption{Evolution of equation (\ref{NL}) at times $t=0$ (black lines), $t=1$ (blue) \& $t=5$ (red), for three choices of initial amplitude $(a)$ $D=0.001$ (linear), $(b)$ $D=0.4$ \& $(c)$ $D=-0.4$. Note we have plotted $- {\rm{v}}  $ in $(c)$ to aid comparison.}
\label{figureappendixB}
\end{center}
\end{figure*}

Consider the isothermal, one-dimensional, nonlinear hydrodynamic equations:
\begin{eqnarray*}
 {\partial \rho \over \partial t}= -  {\partial \over \partial x} \left( \rho {\rm{v}} \right),\;\;  {\partial {\rm{v}} \over \partial t}+ {\rm{v}}  {\partial{\rm{v}}  \over \partial x}=- \frac{1}{\rho}  {\partial p \over \partial x},\;\; p=c_s^2 \rho\;\;.
\end{eqnarray*}
which is equivalent to equation (\ref{MHDequations}) with ${\bf{v}}= {\rm{v}}\hat{\bf{x}}$, ${\bf{B}}={\bf{0}}$. Now let us non-dimensionalise, i.e. ${\rm{v}}={\rm{v}}_0 {\rm{v}}^*$, $\rho={\rho}_0 \rho^*$, $p= p_0 p^*$,   $x = x_0  x^*$, $t={t}_0 t^*$, where we let $*$ denote a dimensionless quantity and ${\rm{v}}_0$, $\rho_0$, $p_0$, $x_0$ and $t_0$ are constants with the dimensions of the variable they are scaling. We set ${\rm{v}}_0 = x_0 / t_0$ and ${\rm{v}}_0 = c_s$, i.e. non-dimensionalise with respect to the sound speed.  For the rest of this appendix, we drop the star indices. Letting $\psi = \log{\rho}$ gives:
\begin{eqnarray*}
 {\partial \psi \over \partial t}+ {\rm{v}}  {\partial \psi \over \partial x}=-  {\partial {\rm{v}} \over \partial x},\;\;  {\partial {\rm{v}} \over \partial t}+  {\rm{v}} {\partial {\rm{v}} \over \partial x}=-  {\partial \psi \over \partial x}.
\end{eqnarray*}
 Adding and subtracting gives:
\begin{eqnarray*}
 {\partial \over \partial t}\left( {\rm{v}}+\psi \right)+ \left({\rm{v}}+1\right)  {\partial \over \partial x} \left( {\rm{v}}+\psi \right)&=&0,\\
 {\partial \over \partial t}\left( {\rm{v}}-\psi \right)+ \left({\rm{v}}-1\right)  {\partial \over \partial x} \left( {\rm{v}}-\psi \right)&=&0.
\end{eqnarray*}
This can be solved in general by the Method of Characteristics:
\begin{eqnarray*}
 {\rm{v}}  &=& \frac{1}{2} F\left[ x- \left( {\rm{v}}  +1\right) t \right] + \frac{1}{2} G\left[ x- \left( {\rm{v}}-1\right) t \right],\\
 \psi &=& \frac{1}{2} F\left[ x- \left( {\rm{v}}+1\right) t \right] - \frac{1}{2} G\left[ x- \left( {\rm{v}}-1\right) t \right],
\end{eqnarray*}
where $F$ and $G$ are arbitrary functions, specified by the initial conditions.

Now let us consider an initial condition in velocity:
\begin{eqnarray}
{\rm{v}}(x) = \left\{  \begin{array}{cl} 
D\left( 1 + x \right)  & {\mathrm{for}}\; {-1 \leq x \leq 0}\\
D\left( 1 - x \right)  & {\mathrm{for}}\; {0 < x \leq 1}\\
0 & { \mathrm{otherwise} }\end{array} \right.\label{NL}\;,
\end{eqnarray}
where $D$ is our initial amplitude. This has a general solution
\begin{eqnarray}
{\rm{v}}(x,t) = \frac{1}{2} F \left[  x- \left({\rm{v}} +1\right) t \right] + \frac{1}{2} F\left[ x- \left({\rm{v}} -1\right) t \right] \label{solution}
\end{eqnarray}
where $F(x)={\rm{v}}(x,0)$. Thus, we can see that the general solution consists of two waves, moving with speeds $ {\rm{v}}  +1$ and $ {\rm{v}}   -1$, which can be thought of as $ {\rm{v}}     +c_s$ and $ {\rm{v}}     -c_s$ due to our choice of non-dimensionalisation.
We can see the evolution of initial condition (\ref{NL}) in Figure \ref{figureappendixB},  at times $t=0$ (black lines), $t=1$ (blue) and $t=5$ (red).  In Figure \ref{figureappendixB}a we see that the velocity pulse (initial amplitude = $D=0.001$) naturally splits into two waves, as expected from the solution, each propagating in opposite directions with amplitude $D / 2$. $  {\rm{v}}     (x,t) = \frac{1}{2} F \left[  x- \left(  {\rm{v}}    +1\right) t \right]$ which, since $  {\rm{v}}   $ is small, can be thought of as $   {\rm{v}}   (x,t) \approx \frac{1}{2} F \left(  x- c_s t \right)$ corresponds to a disturbance travelling in the increasing $x-$direction (to the right), whereas  $      {\rm{v}}      (x,t) = \frac{1}{2} F \left[  x- \left(  {\rm{v}}     -1\right) t \right]\approx \frac{1}{2} F \left(  x+ c_s t \right)$ corresponds to   a disturbance travelling in the decreasing $x-$direction (to the left). This choice of $D$ represents the linear regime.

In Figure \ref{figureappendixB}b we see the evolution of a velocity pulse of initial amplitude = $D=0.4$. Again, the initial pulse splits into two oppositely travelling waves: one propagating to the right (located $\approx x=5$) and one to the left (located $\approx x=-5$).  However, we can now clearly see the nonlinear effects: the two wave pulses are beginning to \lq{tip over}\rq{} and shock, as expected for nonlinear waves. However, non-intuitively, the waves are both developing discontinuities, i.e. shock fronts, on the same faces. The pulse propagating to the right is developing a discontinuity at $x \approx 6$, where the peak is catching up with the the leading footpoint. Conversely, the discontinuity developing in the pulse propagating to the left is located at $x \approx -4$, i.e. where the trailing footpoint is catching up with the peak.

This asymmetry in the development of discontinuities can be explained by equation (\ref{solution}) and can be thought of as follows: when we set our initial condition in velocity, we are effectively  prescribing a background velocity profile, and it is this profile that leads to asymmetry in our system (as opposed to the symmetry present in Figure \ref{figureappendixB}a). A choice of $D>0$ corresponds to a positive background velocity profile that explains the development of the discontinuities on the rightmost faces of both the left and right propagating waves.

This explanation is confirmed by the evolution seen in Figure \ref{figureappendixB}c. Here, we set $D=-0.4$ and have  plotted $-{\rm{v}}$ to aid comparison with $(a)$ and $(b)$. We see that again the initial pulse splits into two oppositely propagating disturbances (each with amplitude $D /2$) and that both these pulses are developing discontinuities on the leftmost faces. This is in agreement with the above interpretation, since a choice of $D<0$  corresponds to a negative background velocity profile.

Interestingly, the footpoints in all three subfigures are located at the same locations. This is because the nonlinear effects are only apparent at large amplitudes and so the footpoints propagate at a speed of unity, i.e. $c_s$.

\end{document}